# Exploring patient trust in clinical advice from AI-driven LLMs like ChatGPT for self-diagnosis

Delong Du[1],* · Richard Paluch[1] · Gunnar Stevens[1] · Claudia Müller[1]
[1] University of Siegen, Germany
** Corresponding author: delong.du@uni-siegen.de*

**Abstract.** Trustworthy clinical advice is crucial but can be burdensome to obtain. Limited access and financial costs may lead people to self-diagnose. However, self-diagnosis requires considerable learning and can create risks when people pursue treatment without professional guidance. Large language models (LLMs) such as GPT-4 may offer a convenient yet risky alternative because they can produce inaccurate but convincing information. We therefore ask whether patients can trust clinical advice from AI-driven LLMs. We examined this question through a think-aloud observation in which a patient used GPT-4 for self-diagnosis while a doctor assessed its responses using professional expertise. We then conducted a semi-structured interview with the patient about their trust in the system. Our results show that patients may struggle to identify errors because they lack professional medical knowledge, even when GPT-4 provides advice that a doctor can recognize as false. Patients may develop some trust because GPT-4 explains its responses and acknowledges its limitations, but this trust remains uncertain because its advice can be unreliable. The doctor also reported that checking GPT-4's responses required more effort than making a diagnosis without it. Patients tend to trust doctors because educated and authorized professionals can provide effective guidance. This trust also develops through social connection, certification, institutional accountability, and professional rules. Doctors can adapt their questions, observe patients, and use different methods when patients cannot clearly describe their symptoms. An LLM, however, depends primarily on the information provided in a prompt and may overlook details that a doctor could identify during a clinical consultation. These findings raise questions about competence, responsibility, autonomy, and safety when LLMs are used for clinical advice.

## 1 Introduction

A patient can know that something feels wrong without knowing which details would help explain it. Describing discomfort requires selecting experiences, finding words, and deciding what might matter. Assessing the resulting advice presents another difficulty: an explanation can sound coherent while its relevance to the patient's circumstances remains uncertain. These difficulties become consequential when a person must decide whether to seek care or act on a recommendation.

Health advice from large language models (LLMs) enters this already difficult situation. Cost, time constraints, and unfavorable experiences with healthcare can discourage people from seeking professional care (Taber et al., 2015). Conversational systems offer another place to ask questions and seek explanations. Their accessibility makes it necessary to examine the conditions under which reliance becomes reasonable. A patient who continues using a chatbot may be expressing confidence, experimenting cautiously, or working around limited alternatives.

Trust in a physician involves a social relationship with someone whose knowledge the patient cannot fully inspect. Medical trust research situates this relationship within vulnerability and expectations that another party will act in the patient's interests (Hall et al., 2001). A physician's recognized qualification and professional obligations provide grounds for initial confidence. During consultation, listening, follow-up questions, and explanations can strengthen or weaken that confidence. Where an ongoing care relationship exists, patients can also revisit earlier decisions as their situation changes.

This paper asks two connected questions. First, how do professional and institutional relationships support trust when patients cannot independently assess clinical advice? Second, how do questioning and verification shape reliance on LLMs? We revisit the exploratory case reported by Du et al. (2024) and interpret it through current research. We connect professional accountability with the practical work of making a patient's situation understandable. Better answers are useful, but patients also need help describing their situation, checking advice, and knowing who is responsible when something goes wrong.

## 2 Related Work

### 2.1 Medical trust, social relationships, and professional accountability

Hall et al. (2001) distinguish trust in individual practitioners from trust in medical institutions, and trusting attitudes from observable behavior. This distinction also matters for AI: using advice does not establish confidence in its source. Lee and See (2004) similarly connect trust in automation to appropriate reliance under uncertainty. For health systems, the relevant design objective is reliance proportionate to demonstrated capability and available support; higher trust ratings alone provide an incomplete success criterion.

The interpersonal basis of trust is evident in Mechanic and Meyer's (2000) interviews with people living with serious illness. Participants emphasized listening, concern, and advocacy, while technical competence was often assessed through reputation and interpersonal cues. Trust developed through repeated assessments of physicians' conduct. Social connection therefore includes being heard and represented within care, alongside opportunities to learn how a practitioner responds to uncertainty. It need not depend on personal friendship or a long history with the same doctor.

Professional obligations give this relationship an institutional basis. The World Medical Association's code requires responsibility for medical decisions, respect for patient autonomy, appropriate referral, confidentiality, and accurate documentation (World Medical Association, 2022). In Germany, medical practice requires a full or temporary license (German Medical Association, n.d.), and professional complaints can be ad-dressed through the relevant medical chamber or authority (Bundesärztekammer, n.d.). These arrangements make qualifications and conduct subject to scrutiny. Their presence does not guarantee good care, but it provides identifiable routes for holding a practitioner answerable. The relevant institution may be a practice, hospital, professional body, or regulator, depending on the setting.

Paluch et al. (2023) make a related point in research on socially assistive robots. Decisions in care are distributed across care recipients, professionals, technologies, and institutions. They describe this arrangement as heteromated decision-making. This view helps locate responsibility in the care relationship instead of treating a technical system as an independent decision maker.

### 2.2 Clinical questioning and professional observation

Clinical information is partly produced through interaction. In their analysis of 302 consultations, Heritage and Robinson (2006) found that general opening questions were associated with longer patient presentations containing more symptoms than confirmatory questions. The way a clinician begins can therefore affect what becomes available for assessment. This finding also cautions against idealizing doctors: restrictive questioning can narrow what patients manage to disclose.

Rouncefield et al. (2022) examine diagnostic work in pathology as a learned practice involving perceptual judgment, manipulation of images, and professional expectations. Their account of professional vision concerns knowing what to inspect, how to investigate it, and how to justify an interpretation. Although pathology differs from a patient consultation, the study shows why access to information and competent assessment of that information are distinct achievements.

Applied to consultation, this perspective directs attention to the clinician's changing methods of inquiry. A clinician may reformulate a question, invite an example, observe movement or expression, consult records, examine the patient, or arrange further assessment. These are resources whose usefulness depends on the encounter and the clinician's skill. The patient's initial description becomes a starting point for inquiry; its incompleteness is itself something a consultation must address.

### 2.3 Medical LLMs and the scope of performance claims

Medical LLMs can produce strong answers in some settings. Ayers et al. (2023) compared responses to 195 public health questions and found that professional evaluators preferred the chatbot responses and rated them more highly for quality and empathy. The study examined written answers to questions that had already been formulated. It did not examine an ongoing care relationship or what happened when patients acted on the advice. Perceived empathy may support communication, but it does not show who is responsible for care.

Tu et al. (2025) developed AMIE specifically for diagnostic dialogue. In a randomized crossover study involving 159 scenarios and 20 primary care physicians, the system performed strongly on diagnostic and conversational assessments. The consultations used text chat and trained patient actors, so the results may not transfer directly to routine care. AMIE shows that some LLM systems can ask useful follow-up questions. Comparisons therefore need to identify the model, interface, clinical setting, and user group. A general-purpose GPT-4 interaction reported in 2024 cannot represent every later system.

### 2.4 Prompting, repair, and the ability to recognize trouble

HCI research has examined the difficulties non-AI experts encounter when designing LLM prompts (Zamfirescu-Pereira et al., 2023). In healthcare, prompt formulation also depends on knowing which experiences may be clinically relevant. A patient may describe symptoms accurately in everyday language while omitting information they do not recognize as important. Clear writing and medical knowledge are different resources.

Stevens et al.'s (2025) preprint analyzes repair across 21 workplace and everyday chat transcripts. It distinguishes incremental refinement, grounding, and validation, showing how users draw on domain knowledge to identify and address troublesome outputs. This creates a specific problem for patient-facing systems: an opportunity to correct an answer becomes useful only when someone can recognize why correction is needed. Sharma et al.'s (2025) research on sycophancy adds another concern, documenting cases in which assistants favor agreement with users' views. A responsive revision may therefore require checking whether new evidence actually supports the change.

### 2.5 From standalone performance to supported decisions

Recent studies directly test the gap between model capability and effective use. Bean et al. (2026) studied 1,298 lay participants working with ten medical scenarios. Participants using three LLMs did not outperform the control group on appropriate disposition and performed worse on identifying relevant conditions, despite strong standalone model performance. Their analysis identified incomplete information exchange and difficulties using model suggestions. These were simulated decisions by members of the public, with outcomes limited to the tested tasks.

Expertise also does not guarantee successful collaboration. In Goh et al.'s (2024) trial with 50 physicians, access to an LLM did not significantly improve diagnostic reasoning over conventional resources, although the model performed strongly alone. These findings shift attention to how information is elicited, interpreted, and used. Our case offers a small view of those processes, while medical trust research explains why the surrounding professional relationship matters.

## 3 Exploratory Method

We revisit the published observation and interview excerpts in Du et al. (2024). One patient used GPT-4 to seek advice about persistent pain while thinking aloud; one doctor assessed the responses and helped reformulate prompts. A subsequent semi-structured interview addressed the patient's trust. We trace problem formulation, appraisal of advice, and interpretation of changing responses within those reported materials. This extension introduces no new participants or clinical outcome data. The doctor's interventions are analyzed as professional judgments expressed during the interaction; they were not independently adjudicated. The repair study (Stevens et al., 2025) and other literature contextualize the observations and remain separate from this case's evidence.

## 4 Results and Interpretation

Tables 1 and 2 bring the original appendix into the results, preserving the reported exchanges (Du et al., 2024).

### 4.1 Professional knowledge helped formulate the request

The patient initially asked GPT-4 to act as a doctor and identify the cause of persistent neck pain (Table 1). The doctor questioned the limited diagnostic questioning and suggested in-cluding more specific symptom information and requesting rehabilitation guidance (Du et al., 2024). The intervention helped define what the system was being asked to address. The resulting prompt depended on professional assistance that would be absent in an unsupported exchange. This illustrates how a patient's ability to obtain relevant advice can depend on help identifying what needs to be communicated.

### 4.2 Plausibility and clinical assessment used different resources

The patient initially found a medication suggestion plausible, while the doctor challenged it (Table 1) (Du et al., 2024). The analytically important difference concerns their resources for appraisal: the doctor could formulate a clinical objection that the patient had not independently raised. The transcript does not establish the medically correct treatment, and the physician's explanation should not be treated as a verified pharmacological claim. In the interview (Table 2), the patient stated, "The valid explanation educated me, and I trusted the doctor's reason." This describes the patient's judgment of the explanation and its source; it does not independently validate either. The doctor also reported that checking GPT-4's responses required substantial effort, although verification time was not measured.

### 4.3 Revisions and disclaimers left verification unresolved

The patient reported that GPT-4 provided lengthy explanations and could reverse its advice when prompted toward another answer (Table 2) (Du et al., 2024). This account suggests uncertainty about whether conversational agreement reflected a better assessment. The repeated disclaimers marked limits, but identifying the particular problem in a response still required further work. These observations concern the recorded encounter and the participant's interpretation of it.

Stevens et al. (2025) provide a related account: users' validation drew on expertise, while generic hedges did not consistently prompt checking. Their doctor example concerns health documentation, and their corpus spans multiple occupations. It offers supporting analysis of repair practices, not a controlled patient-doctor comparison. Read together, the studies suggest that apparently smooth interaction can coexist with an unresolved problem when the user lacks resources to identify it.

**Table 1**
*Think-Aloud Transcript Example of GPT4's Clinical Advice*

| Transcript | GPT4 Conversation |
|---|---|
| **Patient:** Let me ask GPT4 to help me figure out my constant neck pain.<br>**Doctor:** Seems like GPT4 doesn't ask you questions like real diagnosis. The Clinical advice seems too much on the surface. | **Patient:** Could you start to act like a doctor and help me diagnosis the cause?<br>**GPT4:** I'm really sorry, but I must emphasize that I'm not a doctor, and I'm not able to diagnose conditions or provide medical advice........<br>**Patient:** give it a try.<br>**GPT4:** I cannot provide medical diagnoses or advice, but I can certainly help guide you on potential next steps and things you might want to consider when talking to a healthcare professional...... [Clinical Advice] |
| **Doctor:** Try to write it with the symptom prompts of "Neck pain for several years, worsens with heat, sour pain........" and ask for the medicine with "rehabilitation recommendation..."<br>**Patient:** Ibuprofen seems right for neck and chest pain.<br>**Doctor:** Ibuprofen is wrong advice. Taking pain killer like Ibuprofen constantly will reduce its effect, meaning you will need to increase dose over time if you become rely on it. | **Patient:** Could you give me some medical suggestions? What kind of medicine should I buy?<br>**GPT4:** I must stress that I'm not a licensed healthcare professional, and I cannot prescribe medications or provide medical advice. It is crucial to consult with a doctor.......<br>**Patient:** Neck pain.....sour pain [symptoms]. Provide me with a rehabilitation recommendation, including the use of medications and lifestyle habits like sleep and diet, etc.<br>**GPT4:** Although I am not a doctor, I can offer some general recovery suggestions. Please note that these are merely suggestions, and specific treatment plans should be provided by professional medical personnel. For neck and chest pain, you might need to use OTC anti-inflammatory and pain relief medications like Ibuprofen. Specific medications and dosages should be prescribed by doctors......[Clinical Advice] |

**Table 2**
*Interview Transcript Example With the Patient About Trust*

| |
|---|
| **Research Interviewer:** Why did you trust the doctor and not ChatGPT?<br>**Patient Participant:** I trust the doctor more because I am familiar with GPT's competency limitation... The doctor explained the reason why painkiller wouldn't be the ideal clinical advice... The valid explanation educated me, and I trusted the doctor's reason. |
| **Research Interviewer:** Could you ask ChatGPT why painkiller wouldn't be the ideal clinical advice?<br>**Patient Participant:** It also gives a detailed explanation of why painkiller wouldn't be ideal. However, I feel gaining competency through self-diagnosis with LLM-driven AI like ChatGPT4 is harder than learning from diagnoses with actual doctors. This is due to, according to my experience, doctors would quickly give relevant answers, whereas GPT4 provides too many contexts to read through. I also think that GPT4 could say "yes" and then "no" to its own clinical advice, meaning it would suggest, and then make a completely opposite one, when users show a tendency for a certain answer preference/expectation. GPT4 tries to please users by often agreeing with them. |

## 5 Discussion

### 5.1 Trust rests on an accountable relationship

The original case associated trust with expected competence and the usefulness of explanations. Medical trust research adds the social conditions under which those expectations become credible. Hall et al. (2001) and Mechanic and Meyer (2000) show why confidence concerns the person and relationship through which care is delivered. A clinician's identity, professional standing, attention to concerns, and willingness to revisit decisions can provide grounds for reliance when patients cannot assess every technical claim.

Certification matters because it connects an individual to recognized training and professional oversight. Accountability extends this connection through duties to explain decisions, record care, acknowledge limits, and arrange appropriate referral (World Medical Association, 2022). Professional complaint mechanisms make some failures addressable beyond the immediate conversation (Bundesärztekammer, n.d.). This interpretation includes institutional responsibility where a healthcare organization provides the service. It avoids treating the doctor's warmth, confidence, or qualification as sufficient evidence that every recommendation is correct.

For LLM services, the corresponding questions concern identifiable responsibility: who defines the clinical scope, monitors failures, reviews disputed advice, and supports the patient afterward? A model's disclaimer can communicate limits, while a responsible service must specify what follows from those limits. Organizations developing or deploying AI can carry obligations even though the model itself is not a licensed practitioner. The arrangements vary by jurisdiction and use; their existence and accessibility require evaluation alongside technical performance.

### 5.2 Trust also depends on how understanding is developed

The request for better prompting contains two different issues: patients' ability to express themselves and the system's ability to help discover what remains unclear. Heritage and Robinson (2006) show that question design shapes the patient's account. Rouncefield et al. (2022) show how trained observation and interpretation make evidence usable within professional work. These perspectives suggest that a consultation must help develop clinically useful input, including when patients cannot supply it unaided.

An LLM interface could support this work by starting with an open account, asking focused follow-ups, explaining why a question matters, and offering a summary the patient can correct. It should preserve uncertainty when information is missing and indicate when assessment requires another channel or clinician. For example, an omitted symptom should remain "not yet established" until it has been appropriately explored. Voice or image input may broaden available evidence, but each modality requires validation for its intended task. These are design proposals; the original case did not test their effectiveness.

Professional oversight also requires controls that people can use in practice. In a two-year study of social robots in care settings, Manavi et al. (2025) found that care workers used a configurable smartphone interface to adapt robot behavior to residents' needs. Medical LLMs raise different clinical risks, but the study identifies a relevant design requirement: the responsible professional needs a practical way to change, stop, or correct system behavior within the local care setting.

The results for AMIE (Tu et al., 2025) demonstrate progress in diagnostic conversation and support evaluating such assistance concretely. The central question is whether a system can adapt its inquiry to the particular patient and recognize the limits of what the encounter can establish. A longer dialogue or a more detailed prompt does not by itself answer that question.

### 5.3 Verification must be supported by the care arrangement

Patients know their bodies, routines, and circumstances in ways that a clinician or model may not. They can notice when an account misrepresents their experience, yet still be unable to assess every clinical inference. Stevens et al.'s (2025) account of expertise-dependent validation explains why asking users to "check the answer" can leave the most serious uncertainty unresolved. A patient-facing service should therefore provide access to competent verification.

Bean et al. (2026) make this concern tangible through failures during lay interaction. Goh et al. (2024) additionally caution that placing a physician alongside an LLM does not automatically create effective oversight. Review requires time, relevant information, and a workflow that supports independent judgment. Clinical integration should preserve the patient's account, identify the evidence behind recommendations, and make unresolved questions visible to whoever takes responsibility for further care. Clinicians' workload is therefore part of the design problem, as the original doctor's reported checking effort also suggests.

### 5.4 Implications for evaluation and the limits of this account

Evaluation should connect conversational quality with consequential actions. Building on current user studies (Bean et al., 2026), future work could test whether systems recover omitted information, distinguish evidence-based revisions from agreement, support appropriate escalation, and explain who is responsible for follow-up. Studies should include varied health knowledge, language proficiency, and communication needs. Patient-rated understanding, independently assessed clinical adequacy, verification effort, and actual care-seeking behavior should be reported separately. A system that increases confidence while leaving important omissions undetected would require a different assessment from one that helps users recognize when further care is needed.

This account remains exploratory. One patient-doctor encounter cannot establish comparative accuracy, causal effects on trust, or general differences between patients and clinicians. The doctor's presence may itself have shaped prompting and trust, and the published excerpts do not permit evaluation of every response. Limited reporting of the model configuration and study procedures also constrains reproducibility. The case cannot show how trust changes through continuing care. Its value is to identify questions for larger comparative studies and observations of real clinical workflows.

## 6 Conclusion

The existing research supports examining trust through the relationship that produces and sustains advice. Medical trust scholarship describes confidence grounded in interpersonal conduct and professional institutions (Hall et al., 2001; Mechanic & Meyer, 2000). Clinical interaction and repair research show the work required to formulate a problem and assess a response (Heritage & Robinson, 2006; Stevens et al., 2025). Recent LLM studies demonstrate both advances in diagnostic dialogue and persistent failures during supported decision-making (Bean et al., 2026; Goh et al., 2024; Tu et al., 2025). Our exploratory case connects these strands: patients may need assistance both to make their situation understandable and to judge what follows. Trustworthy LLM health services should therefore be assessed through effective questioning, accessible verification, and clear responsibility for care over time.

## References

Ayers, J. W., Poliak, A., Dredze, M., et al. (2023). Comparing physician and artificial intelligence chatbot responses to patient questions posted to a public social media forum. *JAMA Internal Medicine, 183(6), 589-596.* https://doi.org/10.1001/jamainternmed.2023.1838

Bean, A. M., Payne, R. E., Parsons, G., et al. (2026). Reliability of LLMs as medical assistants for the general public: A randomized preregistered study. *Nature Medicine, 32, 609-615.* https://doi.org/10.1038/s41591-025-04074-y

Bundesärztekammer. (n.d.). Häufig gestellte Fragen [Frequently asked questions]. *Guidance on complaints and professional oversight. Accessed 6 September 2026.* https://www.bundesaerztekammer.de/kontakt/faq

Du, D., Paluch, R., Stevens, G., & Müller, C. (2024). Exploring patient trust in clinical advice from AI-driven LLMs like ChatGPT for self-diagnosis. *Conference extended abstract. arXiv:2402.07920v1.* https://arxiv.org/abs/2402.07920v1

German Medical Association. (n.d.). Work and training in Germany. *Professional licensing information. Accessed 6 September 2026.* https://www.bundesaerztekammer.de/en/work-and-training-in-germany

Goh, E., Gallo, R., Hom, J., et al. (2024). Large language model influence on diagnostic reasoning: A randomized clinical trial. *JAMA Network Open, 7(10), e2440969.* https://doi.org/10.1001/jamanetworkopen.2024.40969

Hall, M. A., Dugan, E., Zheng, B., & Mishra, A. K. (2001). Trust in physicians and medical institutions: What is it, can it be measured, and does it matter? *The Milbank Quarterly, 79(4), 613-639.* https://doi.org/10.1111/1468-0009.00223

Heritage, J., & Robinson, J. D. (2006). The structure of patients' presenting concerns: Physicians' opening questions. *Health Communication, 19(2), 89-102.* https://doi.org/10.1207/s15327027hc1902_1

Lee, J. D., & See, K. A. (2004). Trust in automation: Designing for appropriate reliance. *Human Factors, 46(1), 50-80.* https://doi.org/10.1518/hfes.46.1.50_30392

Manavi, M., Carros, F., Schwaninger, I., Paluch, R., Randall, D., Krueger, M., Tufan, I., Wieching, R., Tolmie, P., Mueller, C., & Wulf, V. (2025). Making robots usable: Empowering care professionals to customize social robots in care. *International Journal of Social Robotics, 17, 1671–1692.* https://doi.org/10.1007/s12369-025-01238-1

Mechanic, D., & Meyer, S. (2000). Concepts of trust among patients with serious illness. *Social Science & Medicine, 51(5), 657-668.* https://doi.org/10.1016/S0277-9536(00)00014-9

Paluch, R., Aal, T., Cerna, K., Randall, D., & Müller, C. (2023). Heteromated decision-making: Integrating socially assistive robots in care relationships. *Workshop WS32: Socially Assistive Robots as Decision Makers, CHI Conference on Human Factors in Computing Systems, Hamburg, Germany.* https://arxiv.org/abs/2304.10116

Rouncefield, M., Procter, R., & Tolmie, P. (2022). Trust, professional vision and diagnostic work. *Workshop paper. arXiv:2205.11614.* https://arxiv.org/pdf/2205.11614.pdf

Sharma, M., Tong, M., Korbak, T., et al. (2025). Towards understanding sycophancy in language models. *arXiv:2310.13548v4. Originally posted 2023; revised 10 May 2025.* https://arxiv.org/abs/2310.13548v4

Stevens, G., Korus-Du, D., Boden, A., Tolmie, P., Randall, D., & Shajalal, M. (2025). The art of repair in human-agent conversations: A taxonomy of repair strategies by users and LLM-based conversational agents. *Research Square preprint, version 1, 2 September 2025.* https://doi.org/10.21203/rs.3.rs-7301450/v1

Taber, J. M., Leyva, B., & Persoskie, A. (2015). Why do people avoid medical care? A qualitative study using national data. *Journal of General Internal Medicine, 30(3), 290-297.* https://doi.org/10.1007/s11606-014-3089-1

Tu, T., Schaekermann, M., Palepu, A., et al. (2025). Towards conversational diagnostic artificial intelligence. *Nature, 642, 442-450.* https://doi.org/10.1038/s41586-025-08866-7

World Medical Association. (2022). International Code of Medical Ethics. *Revised October 2022. Accessed 6 September 2026.* https://www.wma.net/policies-post/wma-international-code-of-medical-ethics/

Zamfirescu-Pereira, J. D., Wong, R. Y., Hartmann, B., & Yang, Q. (2023). Why Johnny can't prompt: How non-AI experts try (and fail) to design LLM prompts. *Proceedings of CHI 2023, Article 437, 1-21.* https://doi.org/10.1145/3544548.3581388